\documentclass[a4paper,UKenglish,cleveref, autoref, thm-restate]{lipics-v2021}

\pdfoutput=1
\hideLIPIcs

\title{A 12-CNOT Double Qubit Excitation Gate}

\author{Irfansha Shaik}{Kvantify Aps, Copenhagen, Denmark}{irsh@kvantify.dk}{https://orcid.org/0000-0002-7404-348X}{}

\authorrunning{Irfansha Shaik}

\nolinenumbers

\usepackage{amsmath}
\usepackage{amssymb}
\usepackage{booktabs}
\usepackage{mathtools}
\usepackage{graphicx}
\usepackage{float}
\usepackage{tikz}
\usepackage{quantikz}

\begin{document}

\maketitle

\begin{abstract}
Effective implementation of high-level quantum gates is essential for practical quantum computing.
To the best of our knowledge, we present the first reported 12-CNOT decomposition of the double qubit excitation operator,
improving upon state-of-the-art (SOTA) implementations with 13 CNOTs.
Our new circuit has the lowest CNOT count (12), lowest CNOT depth (8, ${\sim}27\%$ reduction), and lowest total circuit depth (15, $25\%$ reduction) among all the previous SOTA circuits.
Upon allowing output qubit relabeling, the CNOT depth can be further reduced to 7 (a ${\sim}36\%$ reduction from 11).
\end{abstract}

\section{Introduction}
\label{sec:introduction}

Effective implementation of important building blocks (such as high-level quantum gates)
in quantum algorithms is essential for practical quantum computing.
In this work, we look at one such high-level gate, the Double Qubit Excitation Operator,
which implements the equation \eqref{eq:double-excitation} (Eq.~20 of~\cite{yordanov2020efficient}):
\begin{equation}
  \label{eq:double-excitation}
  \begin{split}
    U_{klij}(\theta) = \exp\Big[ -\tfrac{i\theta}{8}\big(
      &\, X_iY_jX_kX_l + Y_iX_jX_kX_l + Y_iY_jY_kX_l + Y_iY_jX_kY_l \\
      &{} - X_iX_jY_kX_l - X_iX_jX_kY_l - Y_iX_jY_kY_l - X_iY_jY_kY_l
    \big)\Big].
  \end{split}
\end{equation}
Using the computational-basis ordering $\lvert q_i q_j q_k q_l\rangle$ in a $4$-qubit system, the operator implements a continuous rotation between the two states $\lvert 0011\rangle$ and $\lvert 1100\rangle$, as shown in \eqref{eq:double-excitation-action}:
\begin{equation}
  \label{eq:double-excitation-action}
  U(\theta)\,\lvert x\rangle =
  \begin{cases}
    \cos\theta\,\lvert 0011\rangle + \sin\theta\,\lvert 1100\rangle, & \lvert x\rangle = \lvert 0011\rangle,\\[2pt]
    \cos\theta\,\lvert 1100\rangle - \sin\theta\,\lvert 0011\rangle, & \lvert x\rangle = \lvert 1100\rangle,\\[2pt]
    \lvert x\rangle, & \text{otherwise.}
  \end{cases}
\end{equation}
The double qubit excitation gate is used as a building block for several quantum algorithms, both near-term and fault-tolerant.
In near-term algorithms, for example, it is used as a CNOT efficient alternative to the
fermionic double excitation operator in variational ground-state ans\"atze such as
unitary coupled cluster (UCCSD)~\cite{peruzzo2014variational,romero2018strategies} in the Variational Quantum Eigensolver (VQE)
and its adaptive variants such as QEB-ADAPT-VQE~\cite{yordanov2021qeb} and FAST-VQE~\cite{majland2023fastvqe}.
In fault-tolerant algorithms, it is used in Trotterized Hamiltonian simulation and
time evolution of the electronic-structure Hamiltonian~\cite{wang2021resource}.
Further, one can use the operator in state preparation,
such as a UCC-type ansatz, for subsequent ground-state energy estimation in fault-tolerant algorithms like
quantum phase estimation (QPE)~\cite{fomichev2024initial}.

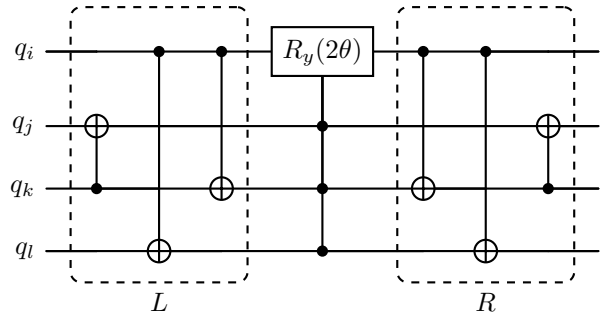
\begin{figure}[H]
  \centering
  \begin{quantikz}
    \lstick{$q_i$} & \qw\gategroup[4,steps=3,style={dashed,rounded corners,inner xsep=2pt},background,label style={label position=below,anchor=north,yshift=-0.2cm}]{$L$} & \ctrl{3} & \ctrl{2} & \gate{R_y(2\theta)} & \ctrl{2}\gategroup[4,steps=3,style={dashed,rounded corners,inner xsep=2pt},background,label style={label position=below,anchor=north,yshift=-0.2cm}]{$R$} & \ctrl{3} & \qw       & \qw \\
    \lstick{$q_j$} & \targ{}   & \qw      & \qw      & \ctrl{-1}           & \qw      & \qw      & \targ{}   & \qw \\
    \lstick{$q_k$} & \ctrl{-1} & \qw      & \targ{}  & \ctrl{-2}           & \targ{}  & \qw      & \ctrl{-1} & \qw \\
    \lstick{$q_l$} & \qw       & \targ{}  & \qw      & \ctrl{-3}           & \qw      & \targ{}  & \qw       & \qw
  \end{quantikz}
  \caption{High-level double-excitation operator.}
  \label{fig:double-excitation-highlevel}
\end{figure}

One can implement the double excitation gate simply using a triple controlled $R_y$ rotation and some CNOT gates.
Figure~\ref{fig:double-excitation-highlevel} shows such a high-level $4$-qubit circuit with triple controlled $R_y(2\theta)$ rotation with $q_i$ as the target qubit,
sandwiched between $L$ and $R$ CNOT circuits.
We refer to Yordanov \emph{et al.}~\cite{yordanov2020efficient} for extended explanation on double qubit excitation operators, which is beyond the scope of this work.
In this paper, we present, to the best of our knowledge, the first reported decomposition of the double-excitation operator with $12$ CNOTs,
improving on the previously reported SOTA implementations with $13$ CNOTs.

The structure of the rest of the paper is as follows.
In Subsection~\ref{subsec:baseline}, we will provide more details on the construction of the high-level circuit, including the role of $L$ and $R$ CNOT circuits.
We also provide a simple 14-CNOT gate implementation using so-called Gray-code expansion~\cite{mottonen2004quantum,shende2006synthesis} of the $C^3R_y(2\theta)$ rotation.
In Subsection~\ref{subsec:previous-sota}, we will present the current SOTA implementations of the double excitation operator with 13-CNOTs.
Finally in Section~\ref{sec:12-cnot-decomposition}, we present a new 12-CNOT circuit for the double excitation operator.
Our new circuit is better in 3 different metrics compared to existing SOTA circuits, i.e., in CNOT count (12), CNOT depth (8, ${\sim}27\%$ reduction; 7 with output qubit relabeling, ${\sim}36\%$ reduction), and circuit depth (15, $25\%$ reduction).

\subsection{A 14-CNOT baseline decomposition using Gray-code expansion}
\label{subsec:baseline}

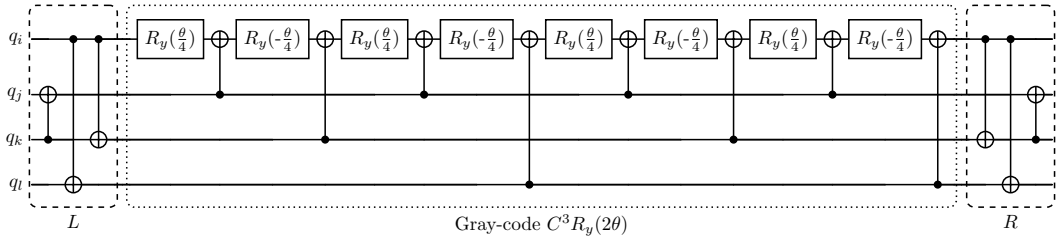
\begin{figure}[ht]
  \centering
  \scalebox{0.72}{%
  \begin{quantikz}[column sep=3.9pt]
    \lstick{$q_i$} & \qw\gategroup[4,steps=3,style={dashed,rounded corners,inner xsep=2pt},background,label style={label position=below,anchor=north,yshift=-0.2cm}]{$L$} & \ctrl{3} & \ctrl{2} & \qw & \qw & \qw & \gate{R_y(\tfrac{\theta}{4})}\gategroup[4,steps=16,style={dotted,rounded corners,inner xsep=2pt},background,label style={label position=below,anchor=north,yshift=-0.2cm}]{Gray-code $C^3R_y(2\theta)$} & \targ{} & \gate{R_y(\text{-}\tfrac{\theta}{4})} & \targ{} & \gate{R_y(\tfrac{\theta}{4})} & \targ{} & \gate{R_y(\text{-}\tfrac{\theta}{4})} & \targ{} & \gate{R_y(\tfrac{\theta}{4})} & \targ{} & \gate{R_y(\text{-}\tfrac{\theta}{4})} & \targ{} & \gate{R_y(\tfrac{\theta}{4})} &\targ{} & \gate{R_y(\text{-}\tfrac{\theta}{4})} & \targ{} & \qw & \qw & \qw & \ctrl{2}\gategroup[4,steps=3,style={dashed,rounded corners,inner xsep=2pt},background,label style={label position=below,anchor=north,yshift=-0.2cm}]{$R$} & \ctrl{3} & \qw & \qw \\
    \lstick{$q_j$} & \targ{}   & \qw      & \qw & \qw & \qw & \qw & \qw & \ctrl{-1} & \qw & \qw       & \qw & \ctrl{-1} & \qw & \qw       & \qw & \ctrl{-1} & \qw & \qw       & \qw & \ctrl{-1} & \qw & \qw & \qw & \qw & \qw & \qw & \qw & \targ{} & \qw  \\
    \lstick{$q_k$} & \ctrl{-1} & \qw      & \targ{} & \qw & \qw & \qw & \qw & \qw       & \qw & \ctrl{-2} & \qw & \qw       & \qw & \qw       & \qw & \qw       & \qw & \ctrl{-2} & \qw & \qw       & \qw & \qw & \qw & \qw & \qw & \targ{} & \qw & \ctrl{-1} & \qw  \\
    \lstick{$q_l$} & \qw       & \targ{}  & \qw & \qw & \qw & \qw & \qw & \qw   & \qw & \qw       & \qw & \qw       & \qw & \ctrl{-3} & \qw & \qw       & \qw & \qw       & \qw & \qw & \qw & \ctrl{-3} & \qw & \qw & \qw & \qw & \targ{} & \qw & \qw
  \end{quantikz}%
  }
  \caption{A 14-CNOT decomposed double-excitation circuit with $C^3R_y(2\theta)$ Gray-code expansion.}
  \label{fig:c3ry-graycode}
\end{figure}

Recall that the double-excitation operator implements 8 Pauli strings, as shown in \eqref{eq:double-excitation}.
One can naively implement the double-excitation operator of 48 CNOTs,
implementing each of the 8 Pauli strings separately using 6 CNOT gates per string (see \S4.7.3 and Fig.~4.19 of~\cite{nielsen2010quantum}).
As discussed earlier, one can also implement the double-excitation operator using a single $C^3R_y(2\theta)$ rotation and some CNOT gates as shown in Figure~\ref{fig:c3ry-graycode}.
Intuitively, we want the controlled rotation to trigger only when the input state is either $\lvert 0011\rangle$ or $\lvert 1100\rangle$.
Precisely, the $L$ CNOT circuit performs this required transformation, as is defined in \eqref{eq:L-network}:
\begin{equation}
  \label{eq:L-network}
  L\,\lvert x\rangle =
  \begin{cases}
    \lvert 0111\rangle, & \lvert x\rangle = \lvert 0011\rangle,\\[2pt]
    \lvert 1111\rangle, & \lvert x\rangle = \lvert 1100\rangle,\\[2pt]
    \lvert x'\rangle \text{ with } q_jq_kq_l \neq 111, & \text{otherwise,}
  \end{cases}
\end{equation}
The $R$ CNOT circuit, on the other hand, performs the inverse transformation of $L$, i.e., $LR = I$.
Depending on the $\theta$, controlled rotation $C^3R_y(2\theta)$ either flips the input basis states $\lvert 0011\rangle$ and $\lvert 1100\rangle$ or leaves them unchanged.
Thus, all the input states excluding $\lvert 0011\rangle$ and $\lvert 1100\rangle$ stay untouched in the output, implementing our desired double-excitation operation.
Decomposition of the triple-controlled $R_y$ rotation is well studied, and multiple 8-CNOT constructions are known.
Here, we use the Gray-code expansion of the $C^3R_y(2\theta)$ rotation~\cite{mottonen2004quantum,shende2006synthesis}, with 8 CNOTs and 8 $R_y$ rotations, resulting in a baseline 14-CNOT double-excitation gate as shown in Figure~\ref{fig:c3ry-graycode}.

\subsection{Previous SOTA 13-CNOT double-excitation gate implementations}
\label{subsec:previous-sota}

In the literature, several 13-CNOT decompositions of the double-excitation operator have been proposed.
The usual approach is to find rewrite rules to absorb some of the CNOTs in the $L$ and $R$ circuits
into the various decompositions of the $C^3R_y(2\theta)$ rotation, resulting in 13-CNOT circuits.
Later in Table~\ref{tab:double-excitation-comparison}, we refer to some well-known examples of these 13-CNOT circuits
of the double-excitation operator, and their metrics.
In this Subsection, we present two of the best 13-CNOT circuits from the literature.
\begin{figure}[ht]
  \centering
  \scalebox{0.61}{%
  \begin{quantikz}[column sep=3pt]
    \lstick{$q_i$} & \ctrl{1} & \qw & \ctrl{2} & \gate{R_y(\tfrac{\theta}{4})} & \ctrl{1} & \gate{R_y(\text{-}\tfrac{\theta}{4})} & \ctrl{3} & \gate{R_y(\tfrac{\theta}{4})} & \ctrl{1} & \gate{R_y(\text{-}\tfrac{\theta}{4})} & \ctrl{2} & \gate{R_y(\tfrac{\theta}{4})} & \ctrl{1} & \gate{R_y(\text{-}\tfrac{\theta}{4})} & \ctrl{3} & \gate{R_y(\tfrac{\theta}{4})} & \ctrl{1} & \gate{R_y(\text{-}\tfrac{\theta}{4})} & \ctrl{2} & \gate{R_z(\tfrac{\pi}{2})} & \qw & \qw & \qw & \ctrl{1} & \qw \\
    \lstick{$q_j$} & \targ{} & \gate{X} & \qw & \gate{H} & \targ{} & \qw & \qw & \qw & \targ{} & \qw & \qw & \qw & \targ{} & \qw & \qw & \qw & \targ{} & \gate{H} & \qw & \gate{X} & \qw & \qw & \qw & \targ{} & \qw \\
    \lstick{$q_k$} & \ctrl{1} & \qw & \targ{} & \qw & \qw & \qw & \qw & \qw & \qw & \gate{H} & \targ{} & \qw & \qw & \qw & \qw & \qw & \qw & \gate{R_z(\text{-}\tfrac{\pi}{2})} & \targ{} & \gate[style={fill=yellow!50}]{H} & \gate{R_z(\text{-}\tfrac{\pi}{2})} & \gate{R_y(\text{-}\tfrac{\pi}{2})} & \gate[style={fill=yellow!50}]{S} & \ctrl{1} & \qw \\
    \lstick{$q_l$} & \targ{} & \gate{X} & \qw & \qw & \qw & \gate{H} & \targ{} & \qw & \qw & \qw & \qw & \qw & \qw & \qw & \targ{} & \gate{H} & \qw & \qw & \qw & \qw & \qw & \gate{X} & \qw & \targ{} & \qw
  \end{quantikz}%
  }
  \caption{A 13-CNOT double-excitation gate by Yordanov et al.~\cite{yordanov2020efficient,yordanov2021qeb} with lowest CNOT depth (11) cf. Table~\ref{tab:double-excitation-comparison}. The two highlighted gates on $q_k$ are added to correct the original circuit.}
  \label{fig:yordanov-2021}
\end{figure}
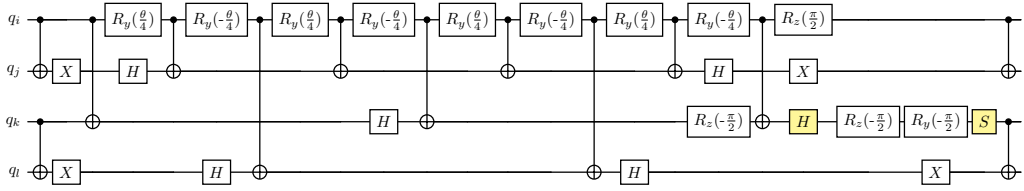
\begin{figure}[ht]
  \centering
  \scalebox{0.66}{%
  \begin{quantikz}[column sep=3pt]
    \lstick{$q_i$} & \gate{S^\dagger} & \ctrl{3} & \ctrl{2} & \ctrl{1} & \gate{H} & \gate{R_z(\tfrac{\theta}{4})} & \targ{} & \gate{R_z(\tfrac{\theta}{4})} & \targ{} & \gate{R_z(\text{-}\tfrac{\theta}{4})} & \targ{} & \gate{R_z(\text{-}\tfrac{\theta}{4})} & \targ{} & \gate{R_z(\tfrac{\theta}{4})} & \targ{} & \gate{R_z(\tfrac{\theta}{4})} & \targ{} & \gate{R_z(\text{-}\tfrac{\theta}{4})} & \targ{} & \gate{R_z(\text{-}\tfrac{\theta}{4})} & \gate{H} & \ctrl{1} & \ctrl{2} & \ctrl{3} & \qw      & \qw \\
    \lstick{$q_j$} & \qw              & \qw      & \qw      & \targ{}  & \qw      & \qw                           & \ctrl{-1} & \qw                         & \qw     & \qw                                   & \ctrl{-1} & \qw                               & \qw     & \qw                           & \ctrl{-1} & \qw                       & \qw     & \qw                                 & \ctrl{-1} & \qw                             & \qw      & \targ{}  & \qw      & \qw      & \qw      & \qw \\
    \lstick{$q_k$} & \qw              & \qw      & \targ{}  & \qw      & \gate{S^\dagger} & \qw                   & \qw     & \qw                           & \qw     & \qw                                   & \qw     & \qw                                   & \ctrl{-2} & \qw                         & \qw     & \qw                       & \qw     & \qw                                 & \qw     & \qw                             & \qw      & \qw      & \targ{}  & \qw      & \gate{S} & \qw \\
    \lstick{$q_l$} & \qw              & \targ{}  & \qw      & \qw      & \qw      & \qw                           & \qw     & \qw                           & \ctrl{-3} & \qw                                 & \qw     & \qw                                   & \qw     & \qw                           & \qw     & \qw                       & \ctrl{-3} & \qw                               & \qw     & \qw                             & \qw      & \qw      & \qw      & \targ{}  & \qw      & \qw
  \end{quantikz}%
  }
  \caption{Nam~\cite{nam2020ground} double-excitation circuit with 13 CNOTs. Among the 13-CNOT reference circuits it has the lowest 1q gate count (11). Cf. Table~\ref{tab:double-excitation-comparison}.}
  \label{fig:nam-2020}
\end{figure}
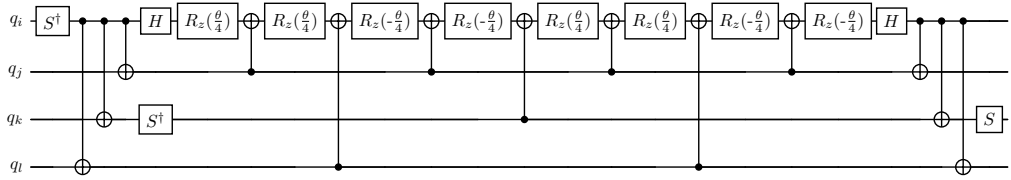
The 13-CNOT circuit by Yordanov \emph{et al.}~\cite{yordanov2020efficient,yordanov2021qeb} as shown in Figure \ref{fig:yordanov-2021}
has the lowest 11 CNOT depth but at the cost of 16 one-qubit (1q) gates.
Figure \ref{fig:nam-2020} on the other hand shows a 13-CNOT circuit by Nam~\cite{nam2020ground}
with the lowest 1q gate count (11) but at the cost of 13 CNOT depth.
Another 13-CNOT circuit by Wang~\cite{wang2021resource} is reported in the
Table~\ref{tab:double-excitation-comparison} with 15 1q gates and 13 CNOT depth.

\section{A 12-CNOT decomposition of the double-excitation operator}
\label{sec:12-cnot-decomposition}

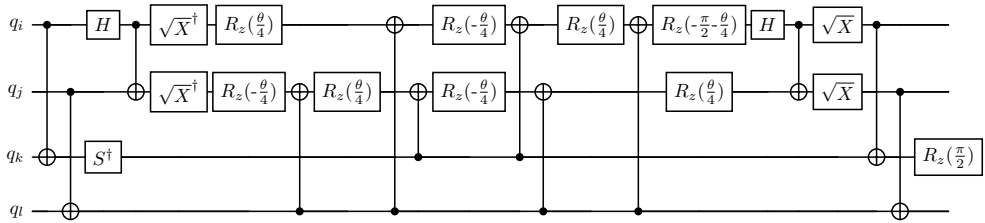
\begin{figure}[!b]
  \centering
  \scalebox{0.72}{%
  \begin{quantikz}[column sep=3pt]
    \lstick{$q_i$} & \ctrl{2} & \qw      & \gate{H} & \ctrl{1} & \gate{\sqrt{X}^{\dagger}} & \gate{R_z(\tfrac{\theta}{4})} & \qw      & \qw                           & \targ{} & \qw      & \gate{R_z(\text{-}\tfrac{\theta}{4})} & \targ{} & \qw      & \gate{R_z(\tfrac{\theta}{4})} & \targ{} & \gate{R_z(\text{-}\tfrac{\pi}{2}\text{-}\tfrac{\theta}{4})} & \gate{H} & \ctrl{1} & \gate{\sqrt{X}} & \ctrl{2} & \qw             & \qw \\
    \lstick{$q_j$} & \qw      & \ctrl{2} & \qw      & \targ{}  & \gate{\sqrt{X}^{\dagger}} & \gate{R_z(\text{-}\tfrac{\theta}{4})} & \targ{}  & \gate{R_z(\tfrac{\theta}{4})} & \qw     & \targ{}  & \gate{R_z(\text{-}\tfrac{\theta}{4})} & \qw     & \targ{}  & \qw     & \qw & \gate{R_z(\tfrac{\theta}{4})}                                                       & \qw      & \targ{}  & \gate{\sqrt{X}} & \qw             & \ctrl{2} & \qw \\
    \lstick{$q_k$} & \targ{}  & \qw      & \gate{S^{\dagger}} & \qw      & \qw              & \qw                           & \qw      & \qw                           & \qw     & \ctrl{-1} & \qw                                   & \ctrl{-2} & \qw    & \qw                           & \qw     & \qw                                                        & \qw      & \qw      & \qw             & \targ{}  & \qw      & \gate{R_z(\tfrac{\pi}{2})} \\
    \lstick{$q_l$} & \qw      & \targ{}  & \qw      & \qw      & \qw                       & \qw                           & \ctrl{-2} & \qw                          & \ctrl{-3} & \qw     & \qw                                   & \qw     & \ctrl{-2} & \qw                        & \ctrl{-3} & \qw                                                      & \qw      & \qw      & \qw             & \qw         & \targ{}  & \qw
  \end{quantikz}%
  }
  \caption{Our best double-excitation circuit: 12 CNOTs, CX-depth 8, 13 1q gates. Cf. Table~\ref{tab:double-excitation-comparison}.}
  \label{fig:our-best}
\end{figure}

\begin{figure}[t]
  \centering
  \scalebox{0.72}{%
  \begin{quantikz}[column sep=3pt]
    \lstick{$q_i$} & \qw               & \ctrl{2} & \qw      & \gate{\sqrt{X}^{\dagger}} & \ctrl{1} & \gate{H}          & \gate{R_z(2\pi\text{-}\tfrac{\theta}{4})} & \targ{}   & \qw      & \gate{R_z(\tfrac{\theta}{4})} & \targ{}   & \qw      & \gate{R_z(2\pi\text{-}\tfrac{\theta}{4})} & \targ{}   & \qw      & \gate{R_z(\tfrac{\theta}{4})}       & \gate{H} & \ctrl{1} & \gate{\sqrt{X}}           & \ctrl{2} & \qw      & \qw                       & \rstick{$q_k$}\qw \\
    \lstick{$q_j$} & \qw               & \qw      & \ctrl{2} & \qw                       & \targ{}  & \gate{\sqrt{X}}   & \gate{R_z(2\pi\text{-}\tfrac{\theta}{4})} & \qw       & \targ{}  & \gate{R_z(\tfrac{\theta}{4})} & \qw       & \targ{}  & \gate{R_z(\pi\text{-}\tfrac{\theta}{4})} & \qw    & \targ{}  & \gate{R_z(\pi\text{+}\tfrac{\theta}{4})} & \qw & \targ{}  & \gate{\sqrt{X}^{\dagger}} & \qw      & \ctrl{2} & \qw                       & \rstick{$q_l$}\qw \\
    \lstick{$q_k$} & \gate{S^{\dagger}} & \targ{}  & \qw      & \qw                       & \qw      & \qw               & \qw                                         & \qw       & \ctrl{-1} & \qw                          & \ctrl{-2} & \qw      & \qw                                         & \qw       & \ctrl{-1} & \qw                                 & \qw      & \qw      & \qw                       & \targ{}  & \qw      & \gate{S}                  & \rstick{$q_i$}\qw \\
    \lstick{$q_l$} & \qw               & \qw      & \targ{}  & \qw                       & \qw      & \qw               & \qw                                   & \ctrl{-3} & \qw      & \qw                           & \qw       & \ctrl{-2} & \qw                                   & \ctrl{-3} & \qw      & \qw                                 & \qw      & \qw      & \qw                       & \qw      & \targ{}  & \qw                       & \rstick{$q_j$}\qw
  \end{quantikz}%
  }
  \caption{A CX-depth-7 variant of our 12-CNOT circuit: 12 CNOTs, CX-depth 7, 13 1q gates, with a final output-wire relabeling $(q_i,q_j,q_k,q_l)\mapsto(q_k,q_l,q_i,q_j)$. Cf. Table~\ref{tab:double-excitation-comparison}.}
  \label{fig:our-best-depth7}
\end{figure}
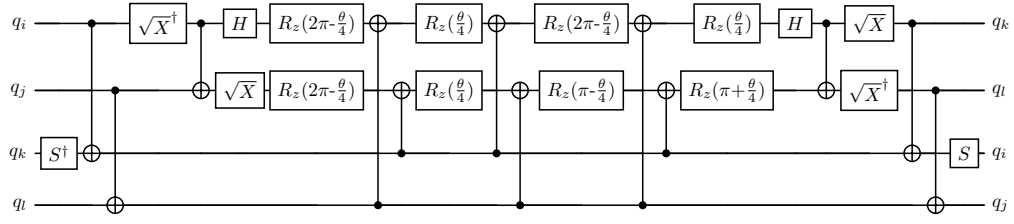

Although several 13-CNOT decompositions of the double-excitation operator have been proposed in the literature,
our literature search found no previously reported implementation with fewer than 13 CNOTs.
In this section, we present the first 12-CNOT decomposition of the double-excitation operator, as shown in Figure~\ref{fig:our-best}.
We have used various circuit synthesis and optimization tools such as Q-Synth~\cite{ShaikvdP2023}, Qiskit Transpiler~\cite{javadiabhari2024qiskit}, Tket~\cite{sivarajah2021tket} to explore different decompositions of the double-excitation operator.
Various synthesis techniques such as Clifford Synthesis~\cite{shaikvdP2025cliffordsynthesis,shaikvdP2026cliffordsynthesisplanning}, CNOT+Rz Synthesis~\cite{ShaikvdP2024cnotsynthesis,li2025hopps}, KAK decomposition~\cite{tucci2005kak} etc. have been useful to extensively explore different decompositions (mainly for excluding dead ends).
Table~\ref{tab:double-excitation-comparison} compares our new 12-CNOT circuit with the previous SOTA 13-CNOT circuits, in $4$ different metrics.
Our new circuit has the lowest CNOT count (12), lowest CNOT depth (8, ${\sim}27\%$ reduction), and lowest total circuit depth (15, $25\%$ reduction) among all the previous SOTA circuits.
Upon allowing output qubit relabeling (as shown in Figure~\ref{fig:our-best-depth7}), the CNOT depth can be further reduced from 8 to 7.
A systematic methodology for exploring different decompositions of the double-excitation operator will be presented in future work.
\begin{table}[b]
  \centering
  \caption{CNOT count, CNOT depth, single-qubit gate count, and total circuit
  depth for double-excitation circuit implementations. Single-qubit gates are
  counted as universal one-qubit (\texttt{u3}) gates, i.e., each maximal run of
  consecutive single-qubit gates on a qubit is merged into one \texttt{u3} gate.}
  \label{tab:double-excitation-comparison}
  \begin{tabular}{lcccc}
    \toprule
    Circuit & CX count & CX depth & 1q gates & Depth \\
    \midrule
    Naive Pauli decomposition~\cite{nielsen2010quantum} & 48 & 48 & 32 & 63 \\
    Baseline: with Gray-code (Fig.~\ref{fig:c3ry-graycode})~\cite{mottonen2004quantum,shende2006synthesis} & 14 & 12 & \textbf{8} & 20 \\
    \cmidrule(lr){1-5}
    PennyLane 2021~\cite{anselmetti2021local,pennylane_doubleexcitation}             & 14 & 12 & 14 & 20 \\
    Wang 2021~\cite{wang2021resource}                                                 & 13 & 13 & 15 & 22 \\
    Nam 2020~\cite{nam2020ground}                                                     & 13 & 13 & 11 & 22 \\
    Yordanov 2020~\cite{yordanov2020efficient,yordanov2021qeb}                        & 13 & 11 & 16 & 20 \\
    \cmidrule(lr){1-5}
    \textbf{This work} (Fig.~\ref{fig:our-best})                                      & \textbf{12} & 8 & 13 & \textbf{15} \\
    \textbf{This work}, output relabeling (Fig.~\ref{fig:our-best-depth7})            & \textbf{12} & \textbf{7} & 13 & \textbf{15} \\
    \bottomrule
  \end{tabular}
\end{table}

\section{Conclusion}
\label{sec:conclusion}

In this work, we presented, to the best of our knowledge, the first reported 12-CNOT decomposition of the double qubit excitation operator.
We compared our new circuit with the previous SOTA 13-CNOT circuits in 4 different metrics.
Our new circuit has the lowest CNOT count (12), lowest CNOT depth (8, ${\sim}27\%$ reduction; 7 with output qubit relabeling, ${\sim}36\%$ reduction), and lowest total circuit depth (15, $25\%$ reduction) among all the previous SOTA circuits.
Further, we only added 2 extra 1q gates (from 11 to 13) compared to the best of the SOTA circuits.
As the double qubit excitation operator can be used as a building block hundreds or thousands of times in practical quantum algorithms, any reduction in such primitives compounds over the full circuit, resulting in significant overall resource savings.

\section*{Acknowledgements}

Author would like to thank Søren Fuglede Jørgensen for the introduction to the problem and the discussions on the topic.
Further, author would also like to thank Jaco van de Pol for the corrections on Yordanov's original circuit in Figure~\ref{fig:yordanov-2021}.
LLMs have been used, mainly GPT 5.6 SOL and Opus 4.8, for setting up experiments, brainstorming, generating Latex figures and Tables, and minor editorial tasks.
Finally, author would like to thank Jaco van de Pol, Patrick Ettenhuber and Asbjørn Frost Teilmann for feedback and correctness checks.
This work was funded by the Innovation Fund Denmark (Grand Solutions) - grant no. 5366-00005B.
\bibliography{references}

@book{nielsen2010quantum,
  author    = {Nielsen, Michael A. and Chuang, Isaac L.},
  title     = {Quantum Computation and Quantum Information: 10th Anniversary Edition},
  publisher = {Cambridge University Press},
  year      = {2010},
  edition   = {10th Anniversary},
  doi       = {10.1017/CBO9780511976667},
}

@article{yordanov2020efficient,
  author  = {Yordanov, Yordan S. and Arvidsson-Shukur, David R. M. and Barnes, Crispin H. W.},
  title   = {Efficient quantum circuits for quantum computational chemistry},
  journal = {Physical Review A},
  volume  = {102},
  number  = {6},
  pages   = {062612},
  year    = {2020},
  doi     = {10.1103/PhysRevA.102.062612},
  eprint  = {2005.14475},
  archivePrefix = {arXiv},
  primaryClass  = {quant-ph},
}

@article{yordanov2021qeb,
  author  = {Yordanov, Yordan S. and Armaos, V. and Barnes, Crispin H. W. and Arvidsson-Shukur, David R. M.},
  title   = {Qubit-excitation-based adaptive variational quantum eigensolver},
  journal = {Communications Physics},
  volume  = {4},
  number  = {1},
  pages   = {228},
  year    = {2021},
  doi     = {10.1038/s42005-021-00730-0},
}

@article{nam2020ground,
  author  = {Nam, Yunseong and Chen, Jwo-Sy and Pisenti, Neal C. and Wright, Kenneth and Delaney, Conor and Maslov, Dmitri and Brown, Kenneth R. and Allen, Stewart and Amini, Jason M. and Apisdorf, Joel and Beck, Kristin M. and Blinov, Aleksey and Chaplin, Vandiver and Chmielewski, Mika and Collins, Coleman and Debnath, Shantanu and Hudek, Kai M. and Ducore, Andrew M. and Keesan, Matthew and Kreikemeier, Sarah M. and Mizrahi, Jonathan and Solomon, Phil and Williams, Mike and Wong-Campos, Jaime David and Moehring, David and Monroe, Christopher and Kim, Jungsang},
  title   = {Ground-state energy estimation of the water molecule on a trapped-ion quantum computer},
  journal = {npj Quantum Information},
  volume  = {6},
  number  = {1},
  pages   = {33},
  year    = {2020},
  doi     = {10.1038/s41534-020-0259-3},
  eprint  = {1902.10171},
  archivePrefix = {arXiv},
  primaryClass  = {quant-ph},
}

@article{wang2021resource,
  author  = {Wang, Qingfeng and Li, Ming and Monroe, Christopher and Nam, Yunseong},
  title   = {Resource-Optimized Fermionic Local-Hamiltonian Simulation on a Quantum Computer for Quantum Chemistry},
  journal = {Quantum},
  volume  = {5},
  pages   = {509},
  year    = {2021},
  doi     = {10.22331/q-2021-07-26-509},
  eprint  = {2004.04151},
  archivePrefix = {arXiv},
  primaryClass  = {quant-ph},
}

@misc{pennylane_doubleexcitation,
  author       = {{Xanadu}},
  title        = {{PennyLane} {\tt qml.DoubleExcitation} operation},
  howpublished = {\url{https://docs.pennylane.ai/en/stable/code/api/pennylane.DoubleExcitation.html}},
  note         = {Accessed 2026-08-06},
  year         = {2024},
}

@article{mottonen2004quantum,
  author  = {M{\"o}tt{\"o}nen, Mikko and Vartiainen, Juha J. and Bergholm, Ville and Salomaa, Martti M.},
  title   = {Quantum circuits for general multiqubit gates},
  journal = {Physical Review Letters},
  volume  = {93},
  number  = {13},
  pages   = {130502},
  year    = {2004},
  doi     = {10.1103/PhysRevLett.93.130502},
  eprint  = {quant-ph/0404089},
  archivePrefix = {arXiv},
}

@article{shende2006synthesis,
  author  = {Shende, Vivek V. and Bullock, Stephen S. and Markov, Igor L.},
  title   = {Synthesis of quantum-logic circuits},
  journal = {IEEE Transactions on Computer-Aided Design of Integrated Circuits and Systems},
  volume  = {25},
  number  = {6},
  pages   = {1000--1010},
  year    = {2006},
  doi     = {10.1109/TCAD.2005.855930},
  eprint  = {quant-ph/0406176},
  archivePrefix = {arXiv},
}

@article{anselmetti2021local,
  author  = {Anselmetti, Gian-Luca R. and Wierichs, David and Gogolin, Christian and Parrish, Robert M.},
  title   = {Local, expressive, quantum-number-preserving {VQE} ans{\"a}tze for fermionic systems},
  journal = {New Journal of Physics},
  volume  = {23},
  number  = {11},
  pages   = {113010},
  year    = {2021},
  doi     = {10.1088/1367-2630/ac2cb3},
  eprint  = {2104.05695},
  archivePrefix = {arXiv},
  primaryClass  = {quant-ph},
}

@article{peruzzo2014variational,
  author  = {Peruzzo, Alberto and McClean, Jarrod and Shadbolt, Peter and Yung, Man-Hong and Zhou, Xiao-Qi and Love, Peter J. and Aspuru-Guzik, Al{\'a}n and O'Brien, Jeremy L.},
  title   = {A variational eigenvalue solver on a photonic quantum processor},
  journal = {Nature Communications},
  volume  = {5},
  pages   = {4213},
  year    = {2014},
  doi     = {10.1038/ncomms5213},
  eprint  = {1304.3061},
  archivePrefix = {arXiv},
  primaryClass  = {quant-ph},
}

@article{romero2018strategies,
  author  = {Romero, Jonathan and Babbush, Ryan and McClean, Jarrod R. and Hempel, Cornelius and Love, Peter J. and Aspuru-Guzik, Al{\'a}n},
  title   = {Strategies for quantum computing molecular energies using the unitary coupled cluster ansatz},
  journal = {Quantum Science and Technology},
  volume  = {4},
  number  = {1},
  pages   = {014008},
  year    = {2018},
  doi     = {10.1088/2058-9565/aad3e4},
  eprint  = {1701.02691},
  archivePrefix = {arXiv},
  primaryClass  = {quant-ph},
}

@article{majland2023fastvqe,
  author  = {Majland, Marco and Ettenhuber, Patrick and Zinner, Nikolaj Thomas},
  title   = {Fermionic Adaptive Sampling Theory for Variational Quantum Eigensolvers},
  journal = {Physical Review A},
  volume  = {108},
  number  = {5},
  pages   = {052422},
  year    = {2023},
  doi     = {10.1103/PhysRevA.108.052422},
  eprint  = {2303.07417},
  archivePrefix = {arXiv},
  primaryClass  = {quant-ph},
}

@article{fomichev2024initial,
  author  = {Fomichev, Stepan and Hejazi, Kasra and Zini, Modjtaba Shokrian and Kiser, Matthew and Fraxanet, Joana and Casares, Pablo Antonio Moreno and Delgado, Alain and Huh, Joonsuk and Voigt, Arne-Christian and Mueller, Jonathan E. and Arrazola, Juan Miguel},
  title   = {Initial state preparation for quantum chemistry on quantum computers},
  journal = {PRX Quantum},
  volume  = {5},
  number  = {4},
  pages   = {040339},
  year    = {2024},
  doi     = {10.1103/PRXQuantum.5.040339},
  eprint  = {2310.18410},
  archivePrefix = {arXiv},
  primaryClass  = {quant-ph},
}

@inproceedings{ShaikvdP2023,
  author       = {Shaik, Irfansha and van de Pol, Jaco},
  title        = {Optimal Layout Synthesis for Quantum Circuits as Classical Planning},
  booktitle    = {2023 IEEE/ACM International Conference on Computer Aided Design (ICCAD)},
  address      = {San Francisco, California, USA},
  organization = {IEEE/ACM},
  year         = {2023},
  doi          = {10.1109/ICCAD57390.2023.10323924},
}

@inproceedings{ShaikvdP2024cnotsynthesis,
  author       = {Shaik, Irfansha and van de Pol, Jaco},
  title        = {Optimal Layout-Aware {CNOT} Circuit Synthesis with Qubit Permutation},
  booktitle    = {27th European Conference on Artificial Intelligence (ECAI 2024)},
  address      = {Santiago de Compostela, Spain},
  publisher    = {IOS Press},
  year         = {2024},
  doi          = {10.3233/FAIA240748},
}

@inproceedings{shaikvdP2025cliffordsynthesis,
  author       = {Shaik, Irfansha and van de Pol, Jaco},
  title        = {{CNOT}-Optimal Clifford Synthesis as {SAT}},
  booktitle    = {28th International Conference on Theory and Applications of Satisfiability Testing (SAT 2025)},
  series       = {LIPIcs},
  volume       = {341},
  pages        = {28:1--28:21},
  publisher    = {Schloss Dagstuhl -- Leibniz-Zentrum f{\"u}r Informatik},
  address      = {Glasgow, Scotland, UK},
  year         = {2025},
  doi          = {10.4230/LIPIcs.SAT.2025.28},
}

@article{shaikvdP2026cliffordsynthesisplanning,
  author  = {Shaik, Irfansha and van de Pol, Jaco},
  title   = {Optimal Clifford Synthesis as Planning},
  journal = {Proceedings of the International Conference on Automated Planning and Scheduling},
  volume  = {36},
  number  = {1},
  pages   = {266--274},
  year    = {2026},
  doi     = {10.1609/icaps.v36i1.42836},
}

@inproceedings{li2025hopps,
  author        = {Li, Xinpeng and Liu, Ji and Xu, Shuai and Hovland, Paul and Chaudhary, Vipin},
  title         = {{HOPPS}: Hardware-Aware Optimal Phase Polynomial Synthesis with Blockwise Optimization for Quantum Circuits},
  booktitle     = {2025 IEEE 32nd International Conference on High Performance Computing, Data, and Analytics (HiPC)},
  year          = {2025},
  doi           = {10.1109/HIPC66333.2025.00010},
  eprint        = {2511.18770},
  archivePrefix = {arXiv},
  primaryClass  = {quant-ph},
}

@misc{javadiabhari2024qiskit,
  author        = {Javadi-Abhari, Ali and Treinish, Matthew and Krsulich, Kevin and Wood, Christopher J. and Lishman, Jake and Gacon, Julien and Martiel, Simon and Nation, Paul D. and Bishop, Lev S. and Cross, Andrew W. and Johnson, Blake R. and Gambetta, Jay M.},
  title         = {Quantum computing with {Qiskit}},
  year          = {2024},
  eprint        = {2405.08810},
  archivePrefix = {arXiv},
  primaryClass  = {quant-ph},
  doi           = {10.48550/arXiv.2405.08810},
}

@article{sivarajah2021tket,
  author  = {Sivarajah, Seyon and Dilkes, Silas and Cowtan, Alexander and Simmons, Will and Edgington, Alec and Duncan, Ross},
  title   = {{t$|$ket$\rangle$}: a retargetable compiler for {NISQ} devices},
  journal = {Quantum Science and Technology},
  volume  = {6},
  number  = {1},
  pages   = {014003},
  year    = {2021},
  doi     = {10.1088/2058-9565/ab8e92},
  eprint  = {2003.10611},
  archivePrefix = {arXiv},
  primaryClass  = {quant-ph},
}

@misc{tucci2005kak,
  author        = {Tucci, Robert R.},
  title         = {An Introduction to {Cartan's} {KAK} Decomposition for {QC} Programmers},
  year          = {2005},
  eprint        = {quant-ph/0507171},
  archivePrefix = {arXiv},
  primaryClass  = {quant-ph},
  doi           = {10.48550/arXiv.quant-ph/0507171},
}

\end{document}